# Betelgeuse at the end of 2019: an historical minimum about to end
Costantino Sigismondi, ICRA/Sapienza , ITIS Ferraris, Rome and AAVSO (SGQ)

**Abstract**
The semi-regular variable star Betelgeuse is undergoing an historical minimum of its brightness. An 8 year series of visual and V-band CCD observations started at the end of 2011 is presented and discussed. Visual methods for comparing magnitudes of angularly distant stars, and performing a differential photometry, needed for such a bright star, are also presented.

**Introduction**
The star named Betelgeuse, the alpha of Orion, is a semi-regular variable star, supergiant.
Its variability was presumably known since the antiquity (Wilk, 1999), but Sir John Herschel is the first to point out its variability in modern time (1840). Allen (1899) reports that in 1852 it began the brightest star of the Northern hemisphere. Helen L. Thomas (1948), conversely, affirmed that stellar variability was not a concern of antiquity.
A [survey of nearly 600](#) visual observations of Betelgeuse was started on [December 2011](#) by the author as SGQ AAVSO observer, and it has been paralleled by Wolfgang Vollman, VOL AAVSO contributor, from Austria, with V-band DSLR CCD differential photometry, with an uncertainty around 6 millimag (4-15 the range).

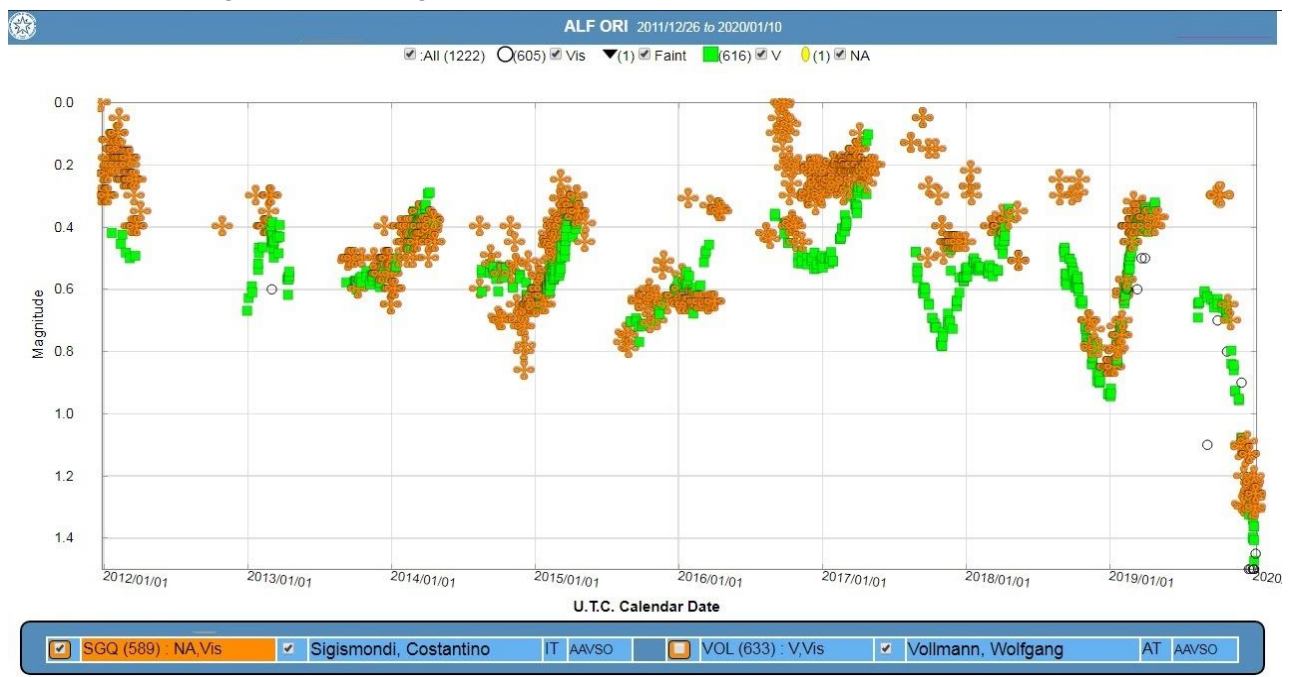

In this figure, obtained by the [AAVSO database](#), the orange crosses are SGQ data, and the green square the V-band VOL data on Betelgeuse from 26/12/2011 to 09/01/2020.
The agreement between SGQ visual data and VOL V-band digital data is generally very good, especially in the years 2013-2016 and 2019. Some departures appear in the beginning of the seasons of observability 2017, 2018 and 2019, when the star is low East in early morning in September 2016, August 2017, September 2018. Conversely good visual observations have been made low West in May 2018 and December 2019-January 2020.

From this plot encompassing 8 whole years is visible the sequence of minima around 1.2 years, especially the three last ones. The present minimum, according to the previous measurements is about to end. This shorter period is consistent with the 425 days one mentioned by Guinan et al. (Atel 13365, 2019) or the 423 days reported in the International Variable Star Index VSX.

For the other period of 5.9 years our database is not long enough to show its evidence strongly, but other wider analyses (Karovska, 1987) are statistically enough strong to show *five statistically significant peaks in the frequency spectrum at 1.05 years, 5.7 years, 6.5 years, 8.8 years, and 20.5 years, indicating multiple periods* (quoted by S. R. Wilk, 1999). Dr. Karovska interpreted the minima as due to mass-ejections.
The previous extremal values reported by Janet Mattei (1999) of 0.4 and 1.3 magnitude, have to be changed to at least 0.1 (8 to 25 april 2017) to 1.3 (end of December 2019).

**Observing Betelgeuse**
To encourage the observations of Betelgeuse I have developed a colored quadrant to perform height measurements, in order to get the airmass through which the star and its comparison stars are observed, and the corresponding dimming in magnitudes.
For Rome I used the relationship magnitude-airmass obtained with the Sun at different altitudes of 0.236 magnitudes/airmass.
The quadrant is in the following figure: at the convergence of the lines a plumb is suspended; the sighting line is the right vertical one from bottom to up.
A web application like stellarium web (set for the observer's location) can be used to evaluate the altitudes at the moment of the observations.

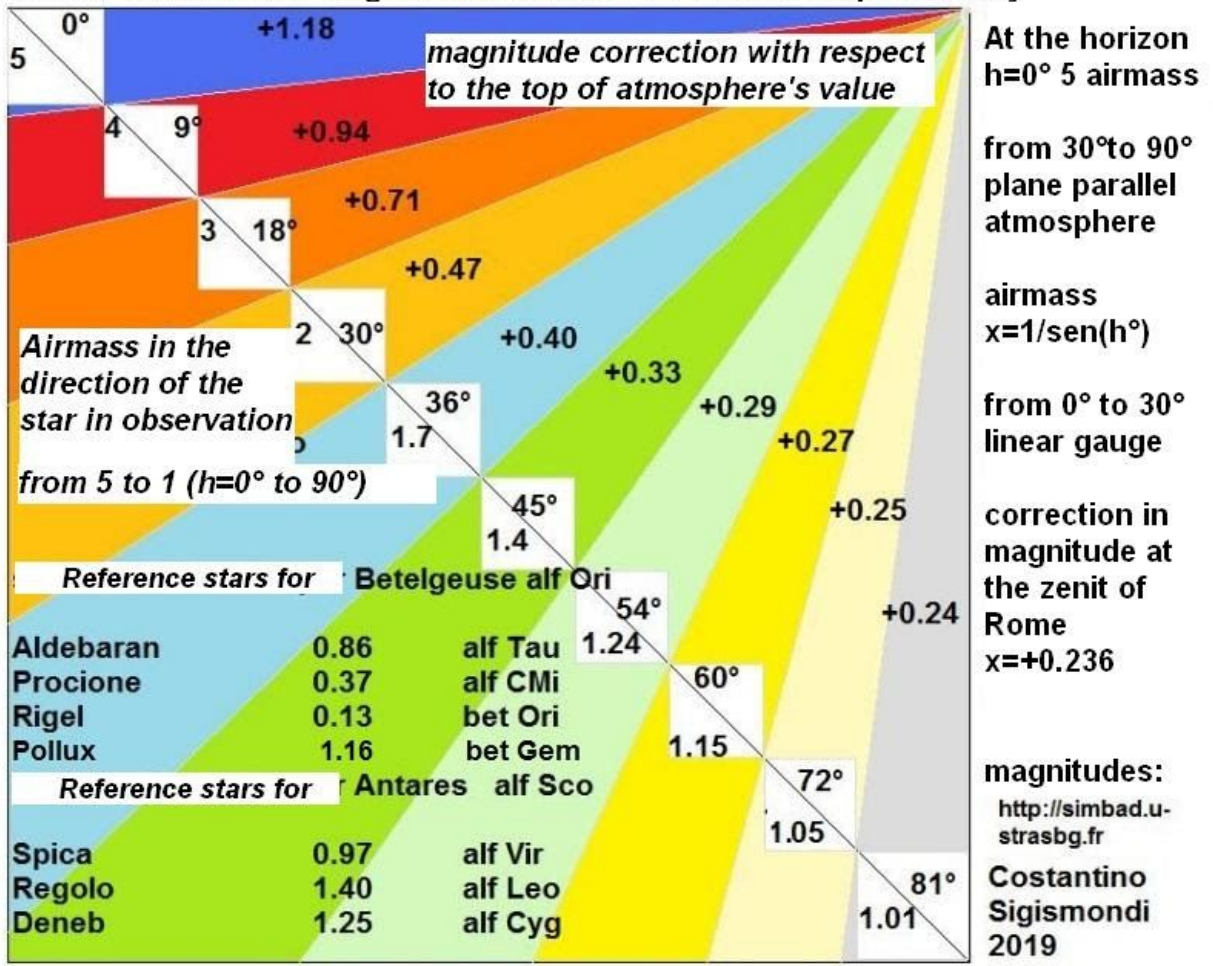

**Purkinje effect**

The human eye in scotopic vision (during the night) is more sensitive to the blue end of the spectrum with respect to the red one. This can be important dealing with red stars, even if Betelgeuse is a very brilliant one. For Betelgeuse and for my 50 years-old eyes, I do not consider induced variations of magnitude, excepted the aforementioned effects at the beginning of the last observational years with the star low East before dusk.

*Visual observers need to be aware of the Purkinje effect and use the quick-glance method to make their observations; do not stare at alf Ori or the comparison stars, as doing so may make them become artificially bright* (Elizabeth O. Waagen AAVSO alert notice 690 (2020)).

**Electronic worksheet for compute the real magnitude**

It is convenient to set up a worksheet to computed the real magnitude after the perceived one and the comparison star data. The order of the columns can be the following

| date | time UTC | bet Gem h[°] | air mass | +delta mag | perceived | alf Ori h[°] | air mass | +delta mag | perceived | real | notes |
|---|---|---|---|---|---|---|---|---|---|---|---|
| 23-dic | 23:58 | 65 | 1,103 | 0,260 | 1,42 | 55 | 1,22 | 0,287 | 1,420 | 1,13 | Argentina |

The airmass is calculated with the cosecant law 1/sen(h°)
The magnitude per airmass is 0.236 for Rome 65 m above sea level.
The perceived magnitude of the reference star is the SIMBAD value + delta mag=airmass x 0.236
The perceived magnitude of Betelgeuse is the one obtained by the various comparisons and the Argelander's method.
The real magnitude is the perceived, minus the delta mag for Betelgeuse's altitude.

**Conclusions**

The opportunity given by this prominent minimum of Betelgeuse occurred in December 2019 to consider the stellar variability to the large public, can be exploited by showing how to perform accurate observations of its photometry, with the unaided eye, up to a few hundredths of magnitude of precision. If the comparisons stars are two, the estimated magnitude of Betelgeuse will be the average of the two estimates, and their semi-difference is a direct estimate of the accuracy of our visual measurement. According to one of the periodicities of Betelgeuse of 1.2 years, we can expect this deep minimum ending, at maximum in February 2020.